\documentclass[twocolumn,showpacs]{revtex4}

\usepackage{graphicx}% Include figure files
\usepackage{dcolumn}% Align table columns on decimal point
\usepackage{bm}% bold math
\usepackage{hyperref}

\begin{document}

%\title{Constructing initial data for binary spinning neutron stars}
\title{Constructing quasi-equilibrium
initial data for binary neutron stars with arbitrary spins}

\author{Wolfgang Tichy}
\affiliation{Department of Physics, Florida Atlantic University,
             Boca Raton, FL  33431, USA}
%\author{Pedro Marronetti, should Pedro be there???}
%\affiliation{Department of Physics, Florida Atlantic University, 
%             Boca Raton, FL 33431, USA}

%\date{$$Id: BNSdata3.tex,v 1.15 2012/10/24 14:06:36 wolf Exp $$}


% Sometimes we want to include preprint numbers, let's put them here
%\preprint{???}

%-------------------------------------------------------------------------
%Useful Definitions
%------------------------------------------------------------------------
%
\newcommand\be{\begin{equation}}
\newcommand\ba{\begin{eqnarray}}

\newcommand\ee{\end{equation}}
\newcommand\ea{\end{eqnarray}}
\newcommand\p{{\partial}}
\newcommand\remove{{{\bf{THIS FIG. OR EQS. COULD BE REMOVED}}}}
%

%-------------------------------------------------------------------------
\begin{abstract}
%-----------------------------------------------------------------------

In general neutron stars in binaries are spinning. Recently, a new
quasi-equilibrium approximation that includes a rotational velocity piece
for each star has been proposed to describe binary neutron stars with
arbitrary rotation states in quasi-circular orbits. We have implemented this
approximation numerically for the first time, to generate initial data for
neutron star binaries with spin. If we choose the rotational velocity piece
such that it equals the Newtonian rigid rotation law, we obtain stars with
fluid 4-velocities that have expansion and shear of approximately zero, as
one would expect for quasi-equilibrium configurations. We also use the new
approach to construct and study initial data sequences for irrotational,
corotating and fixed rotation binaries.

%-----------------------------------------------------------------------
\end{abstract}
%-----------------------------------------------------------------------

\maketitle

%%%%%%%%%%%%%%%%%%%%%%%%%%%%%%%%%%%%%%%%%%%%%%%%%%%%%%%%%%%
\section{Introduction}
%%%%%%%%%%%%%%%%%%%%%%%%%%%%%%%%%%%%%%%%%%%%%%%%%%%%%%%%%%%

Binary neutron stars are at the intersection of two of the most
fascinating topics in astrophysics: Gamma ray bursts and gravitational wave
astronomy. Binary neutron star mergers (together with black hole neutron
star mergers) have been proposed as potential engines for short duration
gamma ray
bursts~\cite{Narayan92,Piran:1999bk,Piran:1999kx,Zhang:2003uk,Piran:2004ba,
Nakar:2007yr,Berger:2010qx,Rezzolla:2011da}. These are likely generated in
the massive accretion disks around the merger remnant: a larger neutron star
or a black hole. In addition, binary neutron star systems are one of the
most promising sources for gravitational wave detectors such as
LIGO~\cite{LIGO:2007kva,LIGO_web},
Virgo~\cite{VIRGO_FAcernese_etal2008,VIRGO_web} or GEO~\cite{GEO_web}.
Several of these detectors have been operating over the last few years,
while several others are in the planning or construction
phase~\cite{Schutz99}. During the inspiral regime, when the two stars are
still well separated they can be well approximated by post-Newtonian theory.
Later, when the stars get close, their matter distributions eventually merge
together to form a single differentially rotating object. Depending on the
total mass, the two progenitors' spins, the equation of state and the
strength of magnetic fields this object can either promptly collapse to a
black hole, or form a hypermassive neutron star. The hypermassive neutron
star is supported against collapse by differential rotation. It can survive
for many dynamical timescales, while angular momentum is gradually
transported from the inner to the outer parts. Eventually the hypermassive
neutron star will also collapse to a black hole surrounded by a massive
torus, that is more massive than in the prompt collapse case. Such systems
could supply the energy required for a short gamma ray
burst~\cite{Narayan92,Piran:1999bk,Piran:1999kx,Zhang:2003uk,Piran:2004ba,
Nakar:2007yr,Rezzolla:2011da}.

In order to make predictions about the last few orbits
and the merger of such systems,
fully non-linear numerical simulations of the Einstein
Equations are required. To start such simulations we need initial
data that describe the binary a few orbits before merger.
The emission of gravitational waves
tends to circularize the orbits~\cite{Peters:1963ux,Peters:1964}.
Thus, during the inspiral, we expect the two neutron stars to be
in quasi-circular orbits around each other with a radius
that shrinks on a timescale much larger than the orbital
timescale. This means that the initial data should
have an approximate helical Killing vector $\xi^{\mu}$.
In general these neutron stars will be spinning. 
In the case of the double pulsar PSR J0737-3039~\cite{Lyne:2004cj}
the spin period of the faster spinning star will be 
$P_{A}(t)=27\mbox{ms}$ at merger~\cite{Tichy:2011gw} and should
thus not be neglected. 
In the quasi-circular regime the orbital timescale will be much shorter than
the spin precession time scale, thus we can assume that the spins are
approximately constant.

To incorporate these ideas and to construct numerical initial data
for binary neutron stars with arbitrary spins and masses
we will use an approach introduced in~\cite{Tichy:2011gw}.
In this approach the stars are given spin by choosing
a rotational velocity for each star. In this way it is possible
to construct stars with both rigid or differential rotation.
In equilibrium we of course expect the stars to be rigidly rotating
such that the expansion and shear of the fluid 4-velocities of each star
vanish. We find that this can be achieved (to good approximation) by setting
the rotational velocity of each star equal to the Newtonian rigid rotation
law.

We also construct and discuss initial data sequences for
irrotational, corotating and fixed rotation binaries.

Spin will have a noticeable effect on the inspiral and merger of the binary if
the spin period is within an order of magnitude of the orbital period. Since
the orbital period is on the order of a few milliseconds during the late
inspiral, we expect interesting spin effects for spin periods on the order
of a few dozen milliseconds or less. To date we have observed only ten
binary neutron star systems, thus it is not clear yet how likely such spin
periods are. Some population synthesis models~\cite{Oslowski:2009zr} suggest
that radio observable pulsars in neutron star binaries do have a
distribution of spin periods that extends down to about 15ms. Other
population synthesis models for pulsars~\cite{Kiel:2008xw} come to similar
findings. However, since such models involve many parameters that are used
to describe sometimes poorly understood physical processes it may be too
early to draw definite conclusions about what spin periods can be expected
in neutron star binaries. One parameter that may be of particular importance
and that illustrates this uncertainty is the magnetic field decay timescale
$\tau_d$. In the models it is assumed that the magnetic field of each star
decays exponentially on this timescale. Since neutron stars spin down due to
magnetic dipole radiation, magnetic field decay can have important effects
for the spin down rate and thus the expected spin periods before merger.
Unfortunately the value of $\tau_d$ is controversial. In the first model
mentioned~\cite{Oslowski:2009zr} the magnetic field decay timescale has to
be $\tau_d \sim 5\mbox{My}$ in order to fit observations, while in the other
model~\cite{Kiel:2008xw} one needs $\tau_d \sim 2000\mbox{My}$.

Throughout we will use units where $G=c=1$.
Latin indices such as $i$ run from 1 to 3 and denote spatial indices,
while Greek indices such as $\mu$ run from 0 to 3 and denote spacetime
indices.
The paper is organized as follows. 
Sec.~\ref{BNSequations} lists the General Relativistic equations
that govern binary spinning neutron stars described by perfect fluids.
In Sec.~\ref{num_method} we briefly describe what algorithm we
use to numerically implement these equations.
We then present some numerical results in Sec.~\ref{results}.
In particular we describe how one should choose the rotational velocity of
each star. We also present particular examples in the form
of several initial data sequences.
We conclude with a discussion of our method in Sec.~\ref{discussion}.
Some derivations are relegated to the appendices~\ref{exp_shear_rot}
and \ref{approx_exp_shear}.

%%%%%%%%%%%%%%%%%%%%%%%%%%%%%%%%%%%%%%%%%%%%%%%%%%%%%%%%%
\section{Binary neutron stars with arbitrary rotation states}
\label{BNSequations}
%%%%%%%%%%%%%%%%%%%%%%%%%%%%%%%%%%%%%%%%%%%%%%%%%%%%%%%%%

In this section we briefly describe the equations governing
binary neutron stars in arbitrary rotation states in General Relativity.
These equations were derived in~\cite{Tichy:2011gw}.

We use the Arnowitt-Deser-Misner (ADM) decomposition of Einstein's
equations and describe the gravitational fields (i.e. the 
4-metric $g_{\mu\nu}$) in terms of the 3-metric
$\gamma_{ij}$, lapse $\alpha$, shift $\beta^i$ 
and the extrinsic curvature $K^{ij}$. We further assume that the
neutron star matter is a perfect fluid with stress-energy tensor
\begin{equation}
T^{\mu\nu} = [\rho_0(1+\epsilon) + P] u^{\mu} u^{\nu} + P g^{\mu\nu}.
\end{equation}
Here $\rho_0$ is the mass density (which is proportional the number
density of baryons), $P$ is the pressure, $\epsilon$ is the internal energy
density divided by $\rho_0$ and $u^{\mu}$ is the 4-velocity of the fluid.
Assuming a polytropic equation of state
\begin{equation}
\label{polytrop}
P = \kappa \rho_0^{1+1/n}
\end{equation}
and defining the specific enthalpy
\begin{equation}
h = 1 + \epsilon + P/\rho_0
\end{equation}
$\rho_0$, $P$ and $\epsilon$ can all be expressed in terms of $h$.
We find
\begin{eqnarray}
\label{rhoPeh_q}
\rho_0   &=& \kappa^{-n} q^n \nonumber \\
P        &=& \kappa^{-n} q^{n+1} \nonumber \\
\epsilon &=& n q ,
\end{eqnarray}
where we have used the abbreviation
\begin{equation}
q = (h-1)/(n+1)
\end{equation}

The fluid 4-velocity $u^{\mu}$ is expressed in terms of the 3-velocity
\begin{equation}
\label{utilde-proj}
^{(3)}\!\tilde{u}^i = h \gamma^i_{\nu}u^{\nu} ,
\end{equation}
which in turn is split into a irrotational piece $D^i \phi$
and a rotational piece $w^i$
\begin{equation}
\label{utilde-split}
^{(3)}\!\tilde{u}^i = D^i \phi + w^i ,
\end{equation}
where $D_i$ is the derivative operator compatible with the 3-metric
$\gamma_{ij}$.

In order to simplify the problem and to obtain elliptic equations we make
several assumptions. The first is the existence of an approximate helical
Killing vector $\xi^{\mu}$, such that
\begin{equation}
\pounds_{\xi} g_{\mu\nu} \approx 0 .
\end{equation}
We also assume similar equations for scalar matter quantities such as $h$.
For a spinning star, however, $\pounds_{\xi} u^{\mu}$ is non-zero.
Instead we assume that 
\begin{equation}
\label{assumption1}
\gamma_i^{\nu} \pounds_{\xi} \left(\nabla_{\nu}\phi\right) \approx 0 ,
\end{equation}
so that the time derivative of the irrotational piece of the
fluid velocity vanishes in corotating coordinates.
We also assume that
\begin{equation}   
\label{assumption2}
\gamma_i^{\nu} \pounds_{\frac{\nabla \phi}{h u^0}} w_{\nu}
\approx 0 ,
\end{equation}
and
\begin{equation}   
\label{assumption3}
^{(3)}\!\pounds_{\frac{w}{h u^0}} w_i \approx 0
\end{equation}
which describe the fact that the rotational piece of the fluid velocity 
is constant along the world line of the star center.

These approximations together with the further assumptions of maximal
slicing
\begin{equation}   
\gamma_{ij} K^{ij} = 0 
\end{equation}
and conformal flatness
\begin{equation}
\label{conflat}
\gamma_{ij} = \psi^4 \delta_{ij} 
\end{equation}
yield the following coupled equations:
\begin{equation}
\label{ham}
\bar{D}^2 \psi 
 + \frac{\psi^5}{32\alpha^2} (\bar{L}B)^{ij}(\bar{L}B)_{ij}
 +2\pi \psi^5 \rho  =  0 ,
\end{equation}
\begin{equation}
\label{mom}   
\bar{D}_j (\bar{L}B)^{ij} 
 -(\bar{L}B)^{ij} \bar{D}_j \ln(\alpha\psi^{-6})  
 -16\pi\alpha\psi^4 j^i  =  0 , 
\end{equation}
\begin{equation}
\label{dt_K_zero}
\bar{D}^2 (\alpha\psi) - \alpha\psi
\left[\frac{7\psi^4}{32\alpha^2}(\bar{L}B)^{ij}(\bar{L}B)_{ij}
      +2\pi\psi^4 (\rho+2S) \right]  = 0 ,
\end{equation}
\begin{equation}
\label{continuity4}
D_i \left[ \frac{\rho_0 \alpha}{h}(D^i \phi + w^i) 
          -\rho_0 \alpha u^0 (\beta^i + \xi^i) \right] = 0 ,
\end{equation}
and
\begin{equation}
\label{h_from_Euler}
h = \sqrt{L^2 - (D_i \phi + w_i)(D^i \phi + w^i)}.
\end{equation}
Here
$(\bar{L}B)^{ij} = \bar{D}^i B^j + \bar{D}^j B^i 
- \frac{2}{3} \delta^{ij} \bar{D}_k B^k$,
$\bar{D}_i = \partial_i$, and we have introduced
\begin{equation}
% \beta^i = B^i + (\Omega\times r)^i
% B^i = \beta^i + \Omega \epsilon^{ij3} (x^j - x_{CM}^j) .
B^i = \beta^i + \xi^i + \Omega \epsilon^{ij3} (x^j - x_{CM}^j) ,
\end{equation}
\begin{eqnarray}
\label{fluid_matter}
\rho   &=& \alpha^2 [\rho_0(1+\epsilon) + P] u^0 u^0 - P, \nonumber \\
j^i    &=& \alpha[\rho_0(1+\epsilon) + P] u^0 u^0
           (u^i/u^0 + \beta^i), \nonumber \\
S^{ij} &=& [\rho_0(1+\epsilon) + P]u^0 u^0 
           (u^i/u^0 + \beta^i) (u^j/u^0 + \beta^j) \nonumber \\
       & &  + P \gamma^{ij} ,
\end{eqnarray}
\begin{eqnarray}
\label{uzero}
u^0 &=& \frac{\sqrt{h^2 + (D_i \phi + w_i)(D^i \phi + w^i)}}{\alpha h},
\nonumber \\
L^2 &=& \frac{b + \sqrt{b^2 - 4\alpha^4 [(D_i \phi + w_i) w^i]^2}}{2\alpha^2},
\nonumber \\
b &=& [ (\xi^i+\beta^i)D_i \phi - C]^2 + 2\alpha^2 (D_i \phi + w_i) w^i ,
\end{eqnarray}
where we sum over repeated spatial indices irrespective of whether they are
up or down and where
$C$ is a constant of integration that in general
has a different value inside each star. Below we will denote these two
values by $C_+$ and $C_-$.
Notice that in an inertial frame the approximate helical 
Killing vector has the components
\begin{equation}
\label{xi_inertial}
\xi^{\mu} = 
\left( 1, -\Omega [x^2 - x^2_{CM}], \Omega [x^1 - x^1_{CM}], 0  \right) .
\end{equation}
Here $x_{CM}^i$ denotes the center of mass position of the system,
and $\Omega$ is the orbital angular velocity, which we have chosen to lie along
the $x^3$-direction.

The elliptic equations (\ref{ham}), (\ref{mom}), (\ref{dt_K_zero})
and (\ref{continuity4}) above have to be solved
subject to the boundary conditions
\begin{equation}
\label{psi_B_alpha_BCs}
\lim_{r\to\infty}\psi = 1, \ \ \ 
\lim_{r\to\infty}B^i = 0, \ \ \
\lim_{r\to\infty}\alpha\psi = 1 
\end{equation}
at spatial infinity, and
\begin{equation}
\label{starBC}
%(\bar{D}^i \phi)\bar{D}_i \rho_0 + \psi^{-2}\bar{w}^i \bar{D}_i \rho_0 
%= h u^0 \psi^4 (\beta^i + \xi^i) \bar{D}_i \rho_0 
(D^i \phi)D_i \rho_0 + w^i D_i \rho_0
= h u^0 (\beta^i + \xi^i) D_i \rho_0
\end{equation}
at each star surface. 
Note that the rotational piece of the fluid velocity $w^i$ can be freely
chosen.

Once the equations (\ref{ham}), (\ref{mom}), (\ref{dt_K_zero}),
(\ref{continuity4}) and (\ref{h_from_Euler}) are solved we know
$h$ (and thus the matter distribution) and the fluid 3-velocity
$^{(3)}\!\tilde{u}^i$ via Eq.~(\ref{utilde-split}).
The 3-metric is obtained from
Eq.~(\ref{conflat}) and the extrinsic curvature is given by
\begin{equation}
K^{ij} =  \frac{1}{2\psi^4 \alpha}(\bar{L}\beta)^{ij} .
\end{equation}

Notice that Eq.~(\ref{h_from_Euler}) can also be written as
\begin{eqnarray}
\label{lnh}
\ln h &+& 
\frac{1}{2}\ln
\left[
\alpha^2 - \left(\beta^i+\xi^i+\frac{w^i}{hu^0}\right)
           \left(\beta_i+\xi_i+\frac{w_i}{hu^0}\right)
\right]
\nonumber \\
&=& -\ln\Gamma + \ln(-C),
\end{eqnarray}
where we have introduced
% % THIS ORIGINAL FORM HAS TYPOS:
% \begin{eqnarray}
% \Gamma = u^0 \sqrt
% {\left[
%  \alpha^2 - \left(\beta^i+\xi^i+\frac{w^i}{hu^0}\right)
%            \left(\beta_i+\xi_i+\frac{w_i}{hu^0}\right)
%  \right]} &&
% \nonumber \\
% \times
% \left(
% 1-\left(\beta^i+\xi^i+\frac{w^i}{hu^0}\right)\frac{D_i\phi}{\alpha^2 hu^0}
%         - \frac{w_i w^i}{(\alpha hu^0)^2}\right) . &&
% \end{eqnarray}
\begin{equation}
\Gamma = 
\frac{
\alpha u^0
\left[
1-\left(\beta^i+\xi^i+\frac{w^i}{hu^0}\right)\frac{D_i\phi}{\alpha^2 hu^0}
        - \frac{w_i w^i}{(\alpha hu^0)^2}\right]
}
{
\sqrt{ 1 -  \left(\beta^i+\xi^i+\frac{w^i}{hu^0}\right)
            \left(\beta_i+\xi_i+\frac{w_i}{hu^0}\right)\frac{1}{\alpha^2} }
} .
\end{equation}

Our approach, while more general, has certain similarities with the approach
in~\cite{Baumgarte:2009fw,Baumgarte:2009fwErr}, hereafter BS. 
For example our
Eq.~(\ref{utilde-split}) reduces to Eq.~(52) of BS if we 
assume $w^i = \eta \xi^i$. 
Furthermore, using the approximations in Eqs.~(\ref{assumption1}),
(\ref{assumption2}) and (\ref{assumption3}), we are able to find the first
integral~(\ref{h_from_Euler}) of the Euler equation. This means that within
our approximations the Euler equation has vanishing curl. That is precisely
what has to hold if the approach in BS (which converts the Euler equation 
into an elliptic equation by taking its divergence) is to succeed. 
The problem with the BS approach is that it simply assumes
$\pounds_{\xi} \tilde{u}^{\mu}=0$ instead of our Eqs.~(\ref{assumption1}),
(\ref{assumption2}) and (\ref{assumption3}). This leads to extra terms in
the Euler equation, which cause a non-vanishing curl.
Thus, the BS approach is not entirely consistent since it
requires zero curl of the Euler equation, while at the same time it uses
an Euler equation with non-vanishing curl.
Furthermore, the boundary condition given by BS for their new elliptic
equation has to be imposed at the star surface. Since the location of the
star surface is another unknown function one needs an equation
or an algorithm to determine this location. No such algorithm is given
by BS. The usual iterative approach where we simply search for the surface
where $h=1$ in each iteration (see e.g. Sec.~\ref{num_method}) will not
work, because the boundary condition given by BS ensures that $h=1$ occurs
at the surface where we impose their boundary condition. So in this kind of
iterative procedure the star surface would always remain at the location of
the initial guess for the surface.

%%%%%%%%%%%%%%%%%%%%%%%%%%%%%%%%%%%%%%%%%%%%%%%%%%%%%%%%%
\section{Numerical method}
\label{num_method}
%%%%%%%%%%%%%%%%%%%%%%%%%%%%%%%%%%%%%%%%%%%%%%%%%%%%%%%%%

To construct initial data we have to solve the elliptic equations
(\ref{ham}), (\ref{mom}), (\ref{dt_K_zero}) and (\ref{continuity4}) together
with the algebraic equation (\ref{h_from_Euler}). This set of equations has
a similar structure as for the well known case of irrotational neutron star
binaries, which has been solved
before~\cite{Bonazzola:1998yq,Gourgoulhon:2000nn,Marronetti:1999ya,
Uryu:1999uu,Marronetti:2003gk,Taniguchi:2002ns,Taniguchi:2003hx,
Uryu:2005vv,Uryu:2009ye}. In this work we will use the SGRID
code~\cite{Tichy:2006qn,Tichy:2009yr,Tichy:2009zr}, which uses
pseudospectral methods to accurately compute spatial derivatives. 
We use the same decomposition into 6 domains as was used
in~\cite{Tichy:2009yr} for the case of corotating neutron star binaries. In
this approach one star center is put on the positive and the other on the
negative $x$-axis. Then complicated coordinate transformations are used to
transform from Cartesian like coordinates $(x,y,z)$ to new coordinates
$(A,B,\varphi)$. Here the coordinates $A$ and $B$ both range from 0 to 1,
and $\varphi$ is a polar angle measured around the $x$-axis. The actual
coordinate transformations are different in each domain. This results in two
domains that cover the outside of each star (including spatial infinity) for
either $x \geq 0$ or $x \leq 0$. These two domains touch at $x=0$. Since
they contain spatial infinity it is trivial to impose the boundary
conditions in Eq.~(\ref{psi_B_alpha_BCs}). The coordinate transformations
contain freely specifiable functions $\sigma_{\pm}(B,\varphi)$ so that one
can always make the inner domain boundaries coincide with the star surfaces.
The inside of each star is covered by two more domains. One of these
stretches from the star surface up to a certain depth inside the star. The
other covers the remainder of the star interior. The elliptic equations
(\ref{ham}), (\ref{mom}) and (\ref{dt_K_zero}) need to be solved in all
domains, while the matter equations~(\ref{continuity4}) and
(\ref{h_from_Euler}) are solved only inside each star. Two of the domain
boundaries always coincide with the neutron star surfaces so that it is
straightforward to impose the boundary condition (\ref{starBC}) for $\phi$
at each star surface. Notice, however, that Eq.~(\ref{continuity4}) and its
boundary condition in Eq.~(\ref{starBC}) do not uniquely specify a solution
$\phi$. If $\phi$ solves both Eqs.~(\ref{continuity4}) and (\ref{starBC})
$\phi + \mbox{const}$ will be a solution as well. In order to obtain a
unique solution we modify the boundary condition by adding the volume
integral $\int_{star}\phi dV$ over the star interior to the left hand side
of Eq.~(\ref{starBC}). Furthermore, in the domains covered by the
$A,B,\varphi$ coordinates we impose the following regularity conditions
along the $x$-axis: \begin{equation} \partial_{\varphi}\Psi = 0 , \ \ \
\partial_{s}\Psi + \partial_{s}\partial_{\varphi}\partial_{\varphi} \Psi = 0
, \end{equation} where $\Psi$ stands for either $\psi$, $B^i$, $\alpha$ or
$\phi$, and $s=\sqrt{y^2 + z^2}$ is the distance from the $x$-axis.

In order to solve the elliptic equations (\ref{ham}), (\ref{mom}),
(\ref{dt_K_zero}) and (\ref{continuity4}) we need a fixed domain
decomposition. However, the location of the star surfaces (where $h=1$) is
not a priorily known, but rather determined by Eq.~(\ref{h_from_Euler}). For
this reason we use the following iterative procedure:
\begin{enumerate}
\item
We first come up with an initial guess for $h$ in 
each star, in practice we simply choose 
Tolman-Oppenheimer-Volkoff solutions 
(see e.g. Chap.~23 in~\cite{Misner73}) for each.
For the irrotational velocity potential we choose
$\phi = \Omega (x_{C*}^1 - x_{CM}^1) x^2$, where $x_{C*}^1$
and $x_{CM}^1$ is the center of the star and the center of mass.
We choose the initial orbital angular velocity according to post-Newtonian
theory.

\item
We then evaluate the residuals [i.e. the $L^2$-norm of the left
hand sides of Eqs.~(\ref{ham}), (\ref{mom}), (\ref{dt_K_zero}) and
(\ref{continuity4})]. If the combined residual
is below a prescribed tolerance we are done and exit the iteration
at this point.

\item
If the residual of Eq.~(\ref{continuity4}) is larger than 
10\% of the combined residuals of Eqs.~(\ref{ham}), (\ref{mom}) and
(\ref{dt_K_zero}), we solve
Eq.~(\ref{continuity4}) for $\phi$. We then reset $\phi$ to
$\phi = 0.2 \phi_{ell} + 0.8 \phi_{old}$, where $\phi_{ell}$ is the just
obtained solution of Eq.~(\ref{continuity4}) and $\phi_{old}$ is the
previous value of $\phi$.

\item
Next we solve the 5 coupled elliptic equations
(\ref{ham}), (\ref{mom}) and (\ref{dt_K_zero}) for 
$\Psi_{ell} = (\psi, B^i, \alpha)_{ell}$. 
We then set $\Psi=(\psi, B^i, \alpha)$ to
$\Psi = 0.4 \Psi_{ell} + 0.6 \Psi_{old}$.

\item
In order to also solve Eq.~(\ref{h_from_Euler}) we need
to know the values of the constants $C_{\pm}$ in each star 
as well as $\Omega$ and $x_{CM}^1$.
We first determine the star centers $x_{C*{\pm}}^1$ by finding the maximum
of the current $h$ along the $x$-axis. Since the location
of each star center is given by $\partial_1 h|_{x_{C*{\pm}}^1} =0$,
Eq.~(\ref{lnh}) [which is equivalent to Eq.~(\ref{h_from_Euler})] yields
\begin{eqnarray}
\label{forcebalance}
\partial_{1} \ln
\left[
\alpha^2 - \left(\beta^i+\xi^i+\frac{w^i}{hu^0}\right)
           \left(\beta_i+\xi_i+\frac{w_i}{hu^0}\right)
\right]\Bigg|_{x_{C*{\pm}}^1} &&
\nonumber \\
= -2\partial_{1}\ln\Gamma\big|_{x_{C*{\pm}}^1} .  &&
\end{eqnarray}
Note that $\beta^i+\xi^i$ is a function of $\Omega$ and $x_{CM}^1$.
We now update $\Omega$ and $x_{CM}^1$ by solving
Eq.~(\ref{forcebalance}) for $\Omega$ and $x_{CM}^1$ so that 
the star centers $x_{C*{\pm}}^1$ remain in the same location, when we update
$h$ according to Eq.~(\ref{h_from_Euler}) or Eq.~(\ref{lnh}).
For this reason Eq.~(\ref{forcebalance}) is sometimes referred 
to as force balance equation. One noteworthy caveat is that
we evaluate the derivative of $\ln\Gamma$ in Eq.~(\ref{forcebalance})
for the $\Omega$ and $x_{CM}^1$ before the update. 
Since we iterate over these steps this approximation does not introduce any
errors. However, it is essential for the overall stability of our
iterative procedure.

\item
Next, we use Eq.~(\ref{h_from_Euler}) to update $h$ in each star, while
at the same time adjusting $C_{\pm}$ such that the rest mass of each star
remains constant. This update is numerically expensive because
the domain boundaries need to be adjusted (by changing
$\sigma_{\pm}(B,\varphi)$) such that they remain
at the star surfaces, which change whenever $h$ is updated.
When we adjust $\sigma_{\pm}(B,\varphi)$ it can be helpful for the stability
of the overall iteration to filter out high frequency
modes in $\sigma_{\pm}(B,\varphi)$ and to impose
$\partial_B \sigma_{\pm}(B,\varphi)|_{B=0,1}=0$. The latter keeps
the stars from drifting away from the $x$-axis during the iterations.

\item
Finally we go back to step 2.
\end{enumerate}

%%%%%%%%%%%%%%%%%%%%%%%%%%%%%%%%%%%%%%%%%%%%%%%%%%%%%
\section{Numerical results}
\label{results}
%%%%%%%%%%%%%%%%%%%%%%%%%%%%%%%%%%%%%%%%%%%%%%%%%%%%%

We have implemented the method described above in the
SGRID code~\cite{Tichy:2006qn,Tichy:2009yr,Tichy:2009zr}.
In this section we present some numerical results using this code.
All our results are presented in units where $G=c=\kappa=1$
and we only use $n=1$ in the polytropic equation of state.

%%%%%%%%%%%%%%%%%%%%%%%%%%%%%%%%%%%%%%%%%%%%%%%%%%%%%
\subsection{Choice of rotational velocity piece $w^i$}

As already mentioned we hold the rest masses fixed during the 
iterations described in Sec.~\ref{num_method}.
Similarly we need to fix the rotational velocity piece
$w^i$ that gives rise to the spin in each star.
Ideally we would like to choose $w^i$ such that the expansion and shear
of the fluid are approximately zero. As we show in Appendices 
\ref{exp_shear_rot} and \ref{approx_exp_shear}, this is true if
the velocity in the corotating frame
$V^{\mu} = u^{\mu}/u^0 - \xi^{\mu}$ is of the form
\begin{equation}
\label{V_is_omega_cross_r}
V^i = \epsilon^{ijk} \hat{\omega}^j [x^k - x_{C*}^k(t)] .
\end{equation}
where $x_{C*}^k$ is the location of the star center, which could be defined
as the point with the highest rest mass density $\rho_0$ or as the center of
mass of the star.
Thus our task is to choose $w^i$ such that Eq.~(\ref{V_is_omega_cross_r})
holds. In~\cite{Tichy:2011gw} we have speculated that a good choice might be
$D_i w^i = 0$, which
can be rewritten in terms of the derivative operator 
$\bar{D}_i = \partial_i$ as $\bar{D}_i \bar{w}^i = 0$ if we introduce
\begin{equation}
\bar{w}^i = \psi^{6} w^i .
\end{equation}  
It is clear that
\begin{equation}
\bar{w}^i = f(|x^n - x_{C*}^n|) \epsilon^{ijk} \omega^j (x^k - x_{C*}^k) ,
\end{equation}
satisfies $\bar{D}_i \bar{w}^i = 0$ for any function $f(|x^n - x_{C*}^n|)$ 
that only depends on the conformal distance from the star's center.
We have tested the simplest case of $f=1$, and find that the
$V^i$ that results from this choice after solving all equations
is similar to Eq.~(\ref{V_is_omega_cross_r}) but with a position
dependent $\hat{\omega}^j$, so that we have differential rotation
in the star which leads to non-zero shear.
Another simple choice is
\begin{equation}
\label{wB_choice}
\bar{w}^i = \psi^6 \epsilon^{ijk} \omega^j (x^k - x_{C*}^k) .
\end{equation}
Numerically, we find that this choice results in a $V^i$ that is very close
to the form in Eq.~(\ref{V_is_omega_cross_r}).
\begin{figure}
\includegraphics[trim = 30 20 30 20,scale=0.87,clip=true]{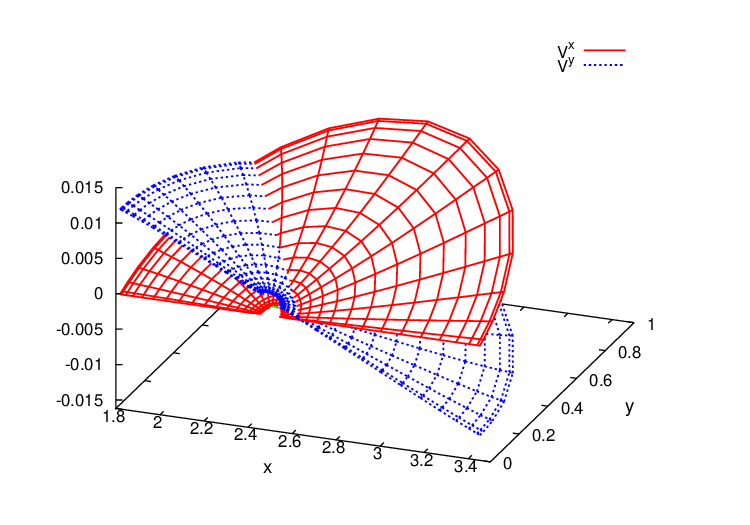}
\caption{\label{m0_1299_1461_oz025P_VR}
The Cartesian $x$- and $y$-components of $V^i$ in
the orbital plane of a the star with rest mass $m_{01} = 0.1461$. The
other star has rest mass $m_{02} = 0.1299$. The separation between both
star centers is $D=5.2885$, which corresponds to an 
orbital angular velocity of $\Omega=0.03928$. Both stars' rotational
velocity piece is given by Eq.~(\ref{wB_choice}) with
$\omega^i=(0,0,0.025)$.
We see that both $V^x$ and $V^y$ vary linearly and thus obey
Eq.~(\ref{V_is_omega_cross_r}).
}
\end{figure}
In Fig.~\ref{m0_1299_1461_oz025P_VR} we show results for an unequal mass
system where both stars have $\bar{w}^i$ as in Eq.~(\ref{wB_choice}) with
$\omega^i=(0,0,0.025)$. Since $V^i$ varies linearly with $x^i$ we
see that it is of the form in Eq.~(\ref{V_is_omega_cross_r}).
The slope of $V^i$ corresponds to 
$\hat{\omega}^i \approx (0,0,-0.015)$, so that
we obtain $\omega^i \approx \Omega^i + \hat{\omega}^i$, which
would hold exactly in Newtonian theory.

%%%%%%%%%%%%%%%%%%%%%%%%%%%%%%%%%%%%%%%%%%%%%%%%%%%%%
\subsection{Initial data sequences}

In order to test our method we have performed simulations with different
rotation states.
\begin{figure}
\includegraphics[scale=0.35,clip=true]{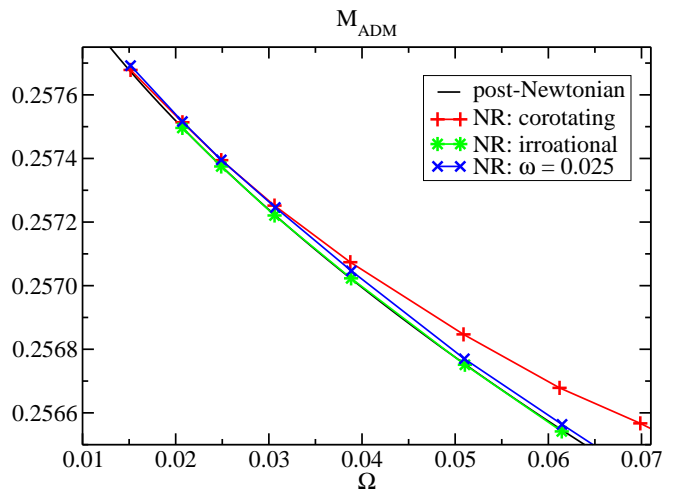}
\caption{\label{M_ADM1299_1461}
The ADM mass for a binary with
rest masses $m_{01}=0.1461$ and $m_{02}=0.1299$
as a function of orbital angular velocity.
Shown are results for post-Newtonian point particles (solid line), 
and three different numerical results (NR) for corotating stars (pluses),
irrotational stars (marked by stars), and a case where both stars
have spin (crosses) with $\omega^i=(0,0,0.025)$.
}
\end{figure}
\begin{figure}
\includegraphics[scale=0.35,clip=true]{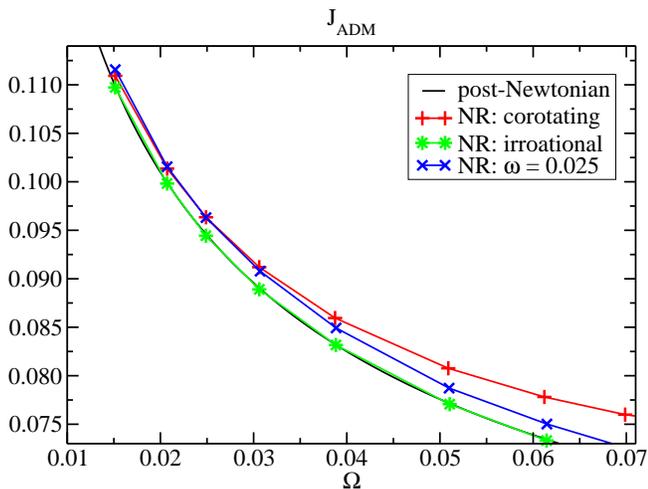}
\caption{\label{J_ADM1299_1461}
The ADM angular momentum for the same binaries as in 
Fig.~\ref{M_ADM1299_1461}.
}
\end{figure}
In Figs.~\ref{M_ADM1299_1461} and \ref{J_ADM1299_1461} we show
how the ADM mass $M_{ADM}$ and the ADM angular momentum
$J_{ADM}$ vary as a function of the orbital angular velocity $\Omega$
for a binaries with rest masses $m_{01}=0.1461$ and $m_{02}=0.1299$.
Note that increasing $\Omega$ corresponds to decreasing separation.
As we can see our numerical results (pluses, stars and crosses)
approach the expected post-Newtonian results
(taken from~\cite{Schaefer93,Tichy02,Tichy03a,Tichy:2003qi}) 
for non-spinning point particles (solid line) for small
$\Omega$. In fact we find that the results for irrotational ($w^i=0$) 
stars (marked by stars in Figs.~\ref{M_ADM1299_1461} and
\ref{J_ADM1299_1461}) agree very well with post-Newtonian 
non-spinning point particle results for all $\Omega$ we have investigated.
On the other hand, for corotating binaries (pluses) we obtain
larger $M_{ADM}$ and $J_{ADM}$ values, especially for larger $\Omega$,
which is expected because to maintain corotation the stars have to spin
more for higher $\Omega$. Finally, we have also investigated the case of a
binary where both stars have the same constant rotational velocity 
$w^i = \epsilon^{ijk} \omega^j (x^k - x_{C*}^k)$ 
with $\omega^i=(0,0,0.025)$. This $\omega^i$ corresponds to a 
spin period of 14ms [for $\kappa=0.018$m$^5$/(kg s$^2$)].
In Figs.~\ref{M_ADM1299_1461} and \ref{J_ADM1299_1461}
this case is denoted by crosses.
Since here the stars always have rotational velocity we obtain
larger $M_{ADM}$ and $J_{ADM}$ values than in the irrotational case for 
all $\Omega$. However, since the rotational velocity is always the same
we get less $M_{ADM}$ and $J_{ADM}$ than in the corotating case (pluses)
for large $\Omega$, and more $M_{ADM}$ and $J_{ADM}$ than in the corotating
case (pluses) for small $\Omega$.

%%%%%%%%%%%%%%%%%%%%%%%%%%%%%%%%%%%%%%%%%%
\section{Discussion}
\label{discussion}
%%%%%%%%%%%%%%%%%%%%%%%%%%%%%%%%%%%%%%%%%%

Realistic neutron stars in binaries are spinning. From observations of
millisecond pulsars we know that these spins can be substantial enough to
influence the late inspiral and merger dynamics of the binary. The spins
might also influence the lifetime of an angular momentum supported
hypermassive neutron star that can form after merger. With more angular
momentum we expect the hypermassive star to survive for longer. This can
have important consequences for both the gravitational waves emitted by the
system and also for the likelihood of a gamma ray burst.

The purpose of this paper is to numerically implement and test a new method
for the computation of binary neutron star initial data with
arbitrary rotation states. This method is derived from the standard
matter equations of perfect fluids together with certain quasi-equilibrium
assumptions. We assume that there is an approximate helical Killing vector
$\xi^{\mu}$ and that Lie derivatives of the metric variables with respect to 
$\xi^{\mu}$ vanish. We also assume that scalar matter variables 
such as $h$ or $\rho_0$ have Lie derivatives that vanish with respect to 
$\xi^{\mu}$. However, since the Lie derivative of the fluid velocity 
$u^{\mu}$ is non-zero for generic spins, we split the fluid velocity
$u^{\mu}$ into an irrotational piece (derived from a potential $\phi$)
and a rotational piece $w^i$, and assume that only
the irrotational piece has a vanishing Lie derivative (see
Eq.~(\ref{assumption1})) with respect to $\xi^{\mu}$. 
Furthermore we know that the spin of each star remains approximately
constant since the viscosity of the stars is insufficient for tidal
coupling~\cite{Bildsten92}. To incorporate this fact,
we use Eqs.~(\ref{assumption2}) and (\ref{assumption3})
which are based on the assumption
that $w^i$ is constant along the star's motion described by the irrotational
velocity piece $\nabla^{\mu} \phi$. 

From these assumptions we obtain the elliptic equations (\ref{ham}),
(\ref{mom}), (\ref{dt_K_zero}) and (\ref{continuity4}) together with the
algebraic equation (\ref{h_from_Euler}). The specific enthalpy $h$ in
Eq.~(\ref{h_from_Euler}) determines the shape of the star surfaces. Since
our elliptic solvers work on a fixed domain decomposition where the star
surfaces have to coincide with domain boundaries we solve this mixture of
elliptic and algebraic equations by iteration. In each iteration we
first solve the elliptic equations for a given $h$ and then use the
algebraic Eq.~(\ref{h_from_Euler}) to update $h$. The stability
of this iterative procedure is improved if we do the following.
(A) 
We typically do not take the $\psi$, $B^i$ and $\alpha$ and
$\phi$   
coming from solving Eqs.~(\ref{ham}),
(\ref{mom}), (\ref{dt_K_zero}) and (\ref{continuity4}) as our new fields.
Rather, we take the average of this solution and 
the values from the previous iteration step
as our new fields. In this way $\psi$, $B^i$ and $\alpha$ and
$\phi$ change less from one iteration step to the next.
(B) 
We use the force balance condition as given in 
Eq.~(\ref{forcebalance}) to update $\Omega$ and $x^1_{CM}$.

For each iteration we also need to specify rotational piece $w^i$ of the
fluid velocity. We have found that the choice in Eq.~(\ref{wB_choice})
works very well in the sense that after numerically
solving all equations it leads to a velocity in the corotating
frame of the form $\vec{V} = \vec{\hat{\omega}} \times \vec{r}$
as in Eq.~(\ref{V_is_omega_cross_r}). 
In Appendices~\ref{exp_shear_rot} and \ref{approx_exp_shear} we show
that this form of $\vec{V}$ results in a fluid 4-velocity with 
an expansion and shear that are approximately zero, as we would expect
for stars in equilibrium.
This means that we have found a simple way to generate initial data
for neutron star binaries that ensure that the star are spinning and
without differential rotation.

We also compare initial data sequences for irrotational, corotating and
fixed rotation binaries. We find that our method yields reasonable results.
All sequences approach post-Newtonian results for large separations. And the
binary sequences with fixed rotation yield higher $M_{ADM}$ and $J_{ADM}$
than corotating configurations for large separations, and lower $M_{ADM}$
and $J_{ADM}$ than corotating configurations for small separations.

\begin{acknowledgments}

It is a pleasure to thank Sebastiano Bernuzzi for helpful discussions.
This work was supported by NSF grants PHY-0855315 and PHY-1204334.

%We also acknowledge TACC at UT Austin for providing
%HPC resources under allocation TG-PHY080019.

\end{acknowledgments}

%%%%%%%%%%%%%%%%%%%%%%%%%%%%%%%%%%%%%%%%%%%%%%%%%%
\appendix
%%%%%%%%%%%%%%%%%%%%%%%%%%%%%%%%%%%%%%%%%%%%%%%%%%

%%%%%%%%%%%%%%%%%%%%%%%%%%%%%%%%%%%%%%%%%%%%%%%%%%
\section{Expansion, shear and rotation of the fluid}
\label{exp_shear_rot}
%%%%%%%%%%%%%%%%%%%%%%%%%%%%%%%%%%%%%%%%%%%%%%%%%%

If $u^{\mu}$ denotes the 4-velocity of the fluid we can split its
derivatives into
\begin{equation}
\nabla_{\nu} u_{\mu} = \frac{1}{3}\Theta P_{\mu\nu} + \sigma_{\mu\nu} + 
\omega_{\mu\nu} - a_{\mu} u_{\nu} ,
\end{equation}
where
\begin{equation}
P_{\mu\nu} = g_{\mu\nu} + u_{\mu} u_{\nu} ,
\end{equation}
and where the expansion, shear, rotation and acceleration are defined as
\begin{equation}
\Theta = P^{\mu\nu} \nabla_{\mu} u_{\nu} ,
\end{equation}
\begin{equation}
\sigma_{\mu\nu} = 
P_{\mu}^{\mu'} P_{\nu}^{\nu'} \nabla_{(\mu'} u_{\nu')}
- \frac{1}{3}\Theta P_{\mu\nu} ,
\end{equation}
\begin{equation}
\omega_{\mu\nu} = P_{\mu}^{\mu'} P_{\nu}^{\nu'} \nabla_{[\nu'} u_{\mu']}
\end{equation}
and
\begin{equation}
a_{\mu} = u^{\nu} \nabla_{\nu} u_{\mu} .
\end{equation}
If the 4-velocity is of the form
\begin{equation}
u^{\mu} = f \bar{u}^{\mu} , 
\end{equation}
where $f$ is any scalar function, it immediately follows 
(from $P_{\mu\nu}u^{\nu}=0$) that
\begin{equation}
\label{Theta_f}
\Theta = f P^{\mu\nu} \nabla_{\mu} \bar{u}_{\nu} ,
\end{equation}
\begin{equation}
\label{sigma_f}
\sigma_{\mu\nu} =
f P_{\mu}^{\mu'} P_{\nu}^{\nu'} \nabla_{(\mu'} \bar{u}_{\nu')}
- \frac{1}{3}\Theta P_{\mu\nu} ,
\end{equation}
and
\begin{equation}
\omega_{\mu\nu} 
= f P_{\mu}^{\mu'} P_{\nu}^{\nu'} \nabla_{[\nu'} \bar{u}_{\mu']} .
\end{equation}

For our purposes it is often convenient to write the 4-velocity as
\begin{equation}
u^{\mu} = u^0 (\xi^{\mu} + V^{\mu}),
\end{equation}
where in an inertial frame the helical Killing vector is given by
Eq.~(\ref{xi_inertial}) and $u^0 = -n_{\mu}u^{\mu}/\alpha$.
Note that $\xi^0=1$ leads to $V^0=0$. From
\begin{equation}
\frac{dx^{\mu}}{dt} = \frac{u^{\mu}}{u^0} = \xi^{\mu} + V^{\mu}
\end{equation}
we see that $V^{\mu}$ can be interpreted as the velocity in the corotating
frame. 

The fact that $\nabla_{(\mu} \xi_{\nu)}=0$ together with 
Eqs~(\ref{Theta_f}) and (\ref{sigma_f}) yields
\begin{equation}
\label{ThetaOfV}
\Theta = u^0 P^{\mu\nu} \nabla_{\mu} V_{\nu} ,
\end{equation}  
and
\begin{equation} 
\label{sigmaOfV}
\sigma_{\mu\nu} =
u^0 P_{\mu}^{\mu'} P_{\nu}^{\nu'} \nabla_{(\mu'} V_{\nu')}
- \frac{1}{3}\Theta P_{\mu\nu} .
\end{equation}
Thus we see that if we have $\nabla_{(\mu} V_{\nu)}\approx 0$, the
expansion and shear are approximately zero.

%%%%%%%%%%%%%%%%%%%%%%%%%%%%%%%%%%%%%%%%%%%%%%%%%%
\section{Expansion and shear if $\vec{V} = \vec{\hat{\omega}} \times \vec{r}$}
\label{approx_exp_shear}
%%%%%%%%%%%%%%%%%%%%%%%%%%%%%%%%%%%%%%%%%%%%%%

Let us assume that $V^{\mu}=(0,V^i)$ is given by
Eq.~(\ref{V_is_omega_cross_r}).
Let us further assume that $V^i$ is small in the sense that
\begin{equation} 
V^i \sim O(\epsilon)
\end{equation}
with $\epsilon \ll 1$. We know (e.g. from
post-Newtonian theory) that the shift is also small. For simplicity
we assume that
\begin{equation} 
\beta^i \sim O(\epsilon).
\end{equation}
Using $\nabla_{(\mu} V_{\nu)} = \pounds_V g_{\mu\nu}/2$ 
and Eq.~(\ref{conflat}) we then find
\begin{eqnarray}
\label{LieVg} 
\nabla_{(0} V_{0)} &=& -V^i \partial_i \alpha + O(\epsilon^2) \nonumber \\
\nabla_{(0} V_{i)} &=& O(\epsilon^2) \nonumber \\
\nabla_{(i} V_{j)} &=& 2\psi^3 V^i \partial_i \psi \delta_{ij} .
\end{eqnarray}
Recall that the Newtonian potential near a mass $m_1$ is given by
\begin{equation}
U \approx \frac{m_1}{r} 
          + \frac{m_2}{D}\left(1+\frac{x^1 - x_{C*}^1(t)}{D}\right) ,
\end{equation}
where $m_2$ is a second mass at a distance $D$ in the $x$-direction.
Since $\psi \propto U$ and $\alpha \propto U$ we find that
\begin{equation}
\label{Vderiv_approx} 
V^i \partial_i \psi \sim V^i \partial_i \alpha \sim O(\epsilon^3) ,
\end{equation}
where we have used the virial theorem and set $\frac{m_2}{D}=O(\epsilon^2)$.

From $u_{\mu} u^{\mu} = -1$ we obtain
\begin{equation} 
u^0 = \frac{1}{\alpha} + O(\epsilon^2) ,
\end{equation}
which leads us to
\begin{equation}
\label{P_approx} 
P^{\mu\nu} = \gamma^{\mu\nu} + O(\epsilon^2) .
\end{equation}
From Eqs~(\ref{ThetaOfV}), (\ref{sigmaOfV}), (\ref{LieVg}), 
(\ref{Vderiv_approx}) and (\ref{P_approx}) we see that
\begin{equation}
\Theta = \sigma_{\mu\nu} = O(\epsilon^3)
\end{equation}
if $V$ is of the form $\vec{\hat{\omega}} \times \vec{r}$ as in 
Eq.~(\ref{V_is_omega_cross_r}). This means that expansion and shear
for this particular $V^i$ are smaller by a factor of $O(\epsilon^2)$
than for a generic $V^i$.

%%%%%%%%%%%%%%%%%%%%%%%%%%%%%%%%%%%%%%%%%%%%%%%

%%%%%%%%%%%%%%%%%%%%%%%%%%%%%%%%%%%%%%%%%%%%%%%
% REFERENCES
%%%%%%%%%%%%%%%%%%%%%%%%%%%%%%%%%%%%%%%%%%%%%%%
\bibliography{references}

\end{document}